\documentclass[%
 reprint,%
 nofootinbib,
 amssymb, amsmath,%
 prc,cha,%
superscriptaddress
]{revtex4-2}
\usepackage[colorlinks=true,linkcolor=blue]{hyperref}%
\usepackage{physics}
\usepackage{graphicx}
\usepackage{lipsum}
\usepackage{mwe}
\usepackage{caption}
\usepackage{cancel}
\usepackage{todonotes}

\usepackage{dcolumn}
\usepackage{bm}
\usepackage{adjustbox}
\usepackage{rotating}
\usepackage{xcolor}
\usepackage{ulem}
\usepackage{braket}

\begin{document}

\title{The $\alpha$-particle condensation in diluted $^{16}\text{O}$ at finite temperature}

\author{M. Davies}
\affiliation{School of Mathematics and Physics, University of Surrey, Guildford, Surrey GU2 7XH, United Kingdom}%

\author{E. Y\"{u}ksel}
\affiliation{School of Mathematics and Physics, University of Surrey, Guildford, Surrey GU2 7XH, United Kingdom}%

\author{J.-P. Ebran}
\affiliation{CEA,DAM,DIF, F-91297 Arpajon, France}
\affiliation{Universit\'e Paris-Saclay, CEA, Laboratoire Mati\`ere en Conditions Extr\^emes, 91680, Bruy\`eres-le-Ch\^atel, France}

\author{E. Khan}
\affiliation{IJCLab, Universit\'e Paris-Saclay, CNRS/IN2P3, 91405 Orsay Cedex, France}
\affiliation{Institut Universitaire de France (IUF)}

\author{P. Stevenson}
\affiliation{School of Mathematics and Physics, University of Surrey, Guildford, Surrey GU2 7XH, United Kingdom}%

\date{\today}

\begin{abstract}
We investigate the effect of temperature on $\alpha$-particle clustering in the diluted nucleus $^{16}\text{O}$ using the multi-constrained finite-temperature relativistic Hartree-Bogoliubov model with the DD-ME2 interaction. At a critical density the nucleus undergoes a Mott-like transition from a homogeneous to a localised configuration characterised by $\alpha$-particle clustering and the emergence of a finite non-axial octupole deformation. We study the interplay between the onset of localisation under nuclear dilution and the suppression of deformation and $\alpha$-particle clustering due to increasing temperature. Investigating the temperature-density plane, our findings indicate that temperature delays the formation of non-axial octupole deformation and $\alpha$-particle clustering in dilute environments. Following the transition from homogeneous to clustered configurations, the non-axial octupole deformation continues to increase with further dilution of the system and becomes nearly independent of temperature.  We found that $\alpha$-particle clusters appear at temperatures up to $T = 4.10$ MeV and at a corresponding normalised density $\rho_{\text{Mott}}/\rho_0 \approx 0.09$.

\end{abstract}

\maketitle


\section{\label{sect:intro}Introduction}

Cluster states, specifically $\alpha$-cluster ones, present a fascinating opportunity to investigate various behaviours displayed within atomic nuclei or nuclear matter. Predictions regarding the occurrence of cluster states in light nuclei date back to 1937, when various possible arrangements of neutrons and protons into distinct groups, such as alpha particles and dineutrons, were proposed by Wheeler \cite{PhysRev.52.1107} and Wefelmeier \cite{Wefelmeier1937-he}. One of the most famous examples of clustering in nuclei is the second $0^{+}$ state in $^{12}\text{C}$ at 7.65 MeV, known as the Hoyle state, which forms from subunits of $\alpha$-particles rather than individual nucleons and is known to be essential for the synthesis of elements
\cite{FREER20141,RevModPhys.89.011002}. There are also theoretical predictions of the sixth $0^+$ state of $^{16}\text{O}$ being an analogue of the Hoyle state, with four $\alpha$-particles instead of three \cite{16O-hoyle-analogue}. Recent experimental studies~\cite{CHEN20231119} have also reported resonances above the 4$\alpha$ separation energy, providing new evidence for the possible existence of a Hoyle-like $\alpha$-clustered structure in ${}^{16}$O. In recent years, many studies have been conducted to better understand the physical mechanisms behind cluster formation in atomic nuclei \cite{PhysRevLett.87.192501,NEFF2004357, 10.1143/PTP.117.655, GIROD-clustering, PhysRevLett.98.032501,Ebran2012,PhysRevC.87.044307, PhysRevC.90.054329, RevModPhys.90.035004, PhysRevC.106.064330, Zhou2023, PRC-16O_QPT,PhysRevC.97.034610,PhysRevC.99.044607,PhysRevC.102.054907,PhysRevC.95.031302,sym16020231} and other environments, such as nuclear matter \cite{ROPKE1982536, ROPKE1983587,PhysRevC.81.015803,PhysRevC.110.L031601,REN2024138463}.

Within the framework of relativistic energy density functional (EDF) theory, an origin of localisation and clustering in atomic nuclei was found to be the depth of the confining potential: a deeper confining potential was shown to support cluster formation in nuclei \cite{Ebran2012,PhysRevC.87.044307, PhysRevC.90.054329, PhysRevC.90.054329}. It was also shown that as the density of a nuclear system decreased, the system undergoes a transition from a homogeneous to a clustered state in both finite nuclei \cite{GIROD-clustering, PRC-16O_QPT} and nuclear matter \cite{ROPKE1982536, ROPKE1983587,PhysRevC.81.015803}, known as a \textit{Mott transition}. In the weak-coupling regime -- characterised by densities near or above the saturation density ($\rho_0 \approx 0.16 \, \text{fm}^{-3}$) -- condensates are dominated by nucleon-nucleon Cooper pairs \cite{nucleon-cooper-pairs1, nucleon-cooper-pairs2}. In contrast, in the strong coupling regime --- characterised by densities much lower than the saturation density --- condensates are proposed to be dominated by four-fermion clusters with zero total spin and isospin, specifically $\alpha$-particles in infinite symmetric nuclear matter \cite{ROPKE1982536, ROPKE1983587, PhysRevC.69.035802, PhysRevC.81.015803, 4f-alpha-condensate}. This implies a phase transition from a homogeneous to clustered configuration through condensation of $\alpha$-particles as the nuclear density progressively decreases. This is also supported by the Hoyle state being much more spatially extended than ground state $^{12}\text{C}$, with its root mean square (rms) radius approximately 1.5 times larger than that of the ground state \cite{hoyle_spatially_extended}. The transition point is governed by a combination of energetic factors and the Pauli exclusion principle, where the clustering recovers the saturation density, lowering the system’s energy compared to a dilute and homogeneous state. The implications of this for finite nuclei have been studied in Refs. \cite{GIROD-clustering, PRC-16O_QPT} within the EDF framework, where the radii are constrained to larger values while imposing a zero quadrupole mass moment to create a dilute nucleus and investigate clustering development. A Mott-like phase transition has been predicted at a critical radius, $R_c$, around $\rho_{\text{Mott}} \sim \rho_0/3$ (where $\rho_{\text{Mott}}$ refers to the Mott density), using both relativistic and non-relativistic EDFs in $N = Z$ nuclei. At this point, the nucleus undergoes an abrupt transition from homogeneous and spherical to a configuration of $\alpha$-particles; for the case of \textsuperscript{16}O this corresponds to a tetrahedral configuration of four $\alpha$-particles.

The properties of nuclei are also influenced by the characteristics of the medium in which they exist. Considering hot nuclear matter in heavy-ion collisions or supernova explosions, temperature also plays a crucial role in describing the nuclear equation of state (EOS) alongside density. Therefore, to fully understand the behaviour of atomic nuclei and the EOS at subsaturation densities, it is crucial to investigate not only the role of density in the formation of cluster states but also the effect of temperature. The occurrence of clustering and condensates in hot and dilute nuclear matter has been investigated in many studies, and it has been found that low-density nuclear matter favours cluster formation at finite temperatures \cite{ROPKE1982536, ROPKE1983587,PhysRevC.79.014002, PhysRevC.79.051301, PhysRevC.81.015803, PhysRevC.97.045805, PhysRevC.110.L031601, REN2024138463}. It has been shown that stable clusters like deuterons, tritons, and $\alpha$ particles can easily form at low temperatures. At high temperatures, clusters dissolve due to increased thermal motion. In Ref. \cite{PhysRevC.106.054309}, the impact of temperature on clustering has also been studied in $^{20}\text{Ne}$, which is known to exhibit pronounced cluster structures in deformed equilibrium shapes at zero temperature \cite{Ebran2012, PhysRevC.87.044307, PhysRevC.90.054329}. It was shown that above the critical temperature of the shape-phase transition, the clustering features disappear \cite{PhysRevC.106.054309}, as temperature weakens the deformation and spreads the density through the surface region, which in turn destroys the localisation features in nuclei.

In this work, we aim to study $\alpha$-particle clustering in $^{16}\text{O}$ under the effects of increasing temperature and decreasing nuclear density. High temperatures, $T\geq 1$ MeV, are known to cause thermal quenching of shell effects, leading to a decrease and eventual disappearance of nuclear deformation in finite nuclei \cite{PhysRevLett.85.26, expanding-limits-nuc-stab, PhysRevC.109.014318}. In this context, we investigate the competition between $\alpha$-particle formation and clustering in nuclei, driven by the dilution of the nucleus and the development of non-axial octupole deformation, alongside the opposing effects of reduced deformation and suppressed $\alpha$-particle clustering as temperature increases. We determine a critical temperature and density at which deformation completely vanishes and clustering disappears.


Our paper is organised as follows. First, in Sec.~\ref{sect:theory}, we provide a brief overview of the relativistic EDF theory realised at the multi-constrained finite temperature relativistic Hartree-Bogoliubov (FT-RHB) level. In Sec.~\ref{sect:results}, we perform multi-constrained FT-RHB calculations at both zero and finite temperatures, where the system's radius is constrained to larger values in order to create a dilute nucleus. By diluting the nucleus, we investigate the formation and behaviour of $\alpha$-clustering in $^{16}\text{O}$ at zero and finite temperatures. The main findings of the paper are summarised in Sec.~\ref{sect:conc}.


\section{\label{sect:theory}Theoretical Formalism} 
At finite temperatures, atomic nuclei are treated as a grand-canonical ensemble in which both heat and particles are exchanged, and this system is characterised by the grand potential \cite{GOODMAN198130, PhysRevC.34.1942,BONCHE1984278}:
\begin{equation}
    \Omega = E - TS - \sum_{q=p,n} \lambda_q N_q,
    \label{eqn:theory-grand_pot}
\end{equation}
where $E$ is the total energy, $T$ is the temperature, and $S$ is the entropy. The chemical potential is specified for both protons ($\lambda_p$) and neutrons ($\lambda_n$) and $N$ is the associated particle number. The quasiparticle (q.p.) operators are defined through a unitary Bogoliubov transformation \cite{ring1980nuclear}:
\begin{equation}
    \alpha^{\dagger}_k = \sum_n U_{nk}c^{\dagger}_n + V_{nk}c_n.
\end{equation}
Here, $c^{\dagger}_n$ and $c_n$ are the single-nucleon creation and annihilation operators, respectively, while $\alpha^{\dagger}_k$ denotes the q.p. creation operator. The $U$ and $V$ represent the Hartree–Bogoliubov wave functions, and the index $n$ corresponds to an oscillator basis
(see Refs. \cite{ring1980nuclear,VRETENAR2005101, dirhb} for detailed information).

To describe the properties of highly-excited or hot nuclei, we use the finite-temperature relativistic Hartree-Bogoliubov (FT-RHB) approach. The FT-RHB equations are found by minimising the grand potential with respect to the density operator, such that $\delta \Omega =0$ \cite{GOODMAN198130}. The FT-RHB equations in the quasiparticle basis are given by
\begin{equation}
    \left(
    \begin{array}{cc}
    h_D-\lambda-m & \Delta \\
    -\Delta^{*} & -h_D^{*}+\lambda +m
    \end{array}
    \right)
    \left(\begin{array}{l}
    U_k \\ V_k
    \end{array}
    \right)
    = E_k \left(\begin{array}{l}
    U_k \\ V_k
    \end{array}
    \right),
    \label{eqn:theory-FTRHB}
\end{equation}
where $h_D$ is the single-nucleon Dirac Hamiltonian, and $\Delta$ is the pairing field. The chemical potential, $\lambda$, is determined by enforcing the conservation of the total particle number on average. The $U_k$ and $V_k$ are the usual q.p. wave functions with energy $E_{k}$. The single-nucleon Dirac Hamiltonian $h_{D}$ is given as:
\begin{equation}
    h_D = -i \bm{\alpha} \bm{\nabla} + V(\bm{r}) + \beta (m + S(\bm{r}))
    \label{eqn:theory-hD}
\end{equation}
where $m$ is the nucleon mass, $S(\bm{r})$ is the scalar potential, and $V(\bm{r})$ is the vector potential, each of which depends on the chosen form of the relativistic EDF (see Ref. \cite{dirhb} for more details). Within the FT-RHB framework, the scalar, vector, and isovector densities are given by \cite{GOODMAN198130, PhysRevC.34.1942}
\begin{align}
    &\rho_s = \sum_{E_k>0} V_k^{\dagger} \gamma^0 (1-f_k) V_k + U_k^T \gamma^0 f_k U_k^{*},\\
    &\rho_v = \sum_{E_k>0} V_k^{\dagger} (1-f_k) V_k + U_k^T f_k U_k^{*},\\
    &\rho_{tv} = \sum_{E_k>0} V_k^{\dagger} \tau_3 (1-f_k) V_k + U_k^T \tau_3 f_k U_k^{*},
\end{align}
where $\tau_{3}$ is the third component of the Pauli isospin matrix and $f_k$ is the Fermi-Dirac factor and given by

\begin{equation}
    f_k = \frac{1}{1+\text{e}^{\beta E_k}}.
    \label{eqn:theory-Fermi_Dirac}
\end{equation}
Here, $E_k$ represents the quasiparticle energy, and $\beta = 1/k_B T$, where $k_B$ is the Boltzmann constant and $T$ is the temperature. In this work, calculations are performed using the meson-exchange DD-ME2 functional \cite{PhysRevC.71.024312}. The pairing field is given by 
\begin{equation}
    \Delta_{ll'} = \frac{1}{2} \sum_{kk'}V_{ll'kk'}^{pp}\kappa_{kk'},
\end{equation}
where $V_{ll'kk'}^{pp}$ is the matrix element of the particle-particle (pairing) force. The pairing tensor is defined as
\begin{equation}
    \kappa = \sum_{E_k>0} V_k^{*} (1-f_k) U_k^T + U_k f_k V_k^{\dagger}.
\end{equation}
 
The particle-particle force is separable in momentum space, and in coordinate space has the form \cite{TIAN200944, PhysRevC.79.064301}
\begin{equation}
    V^{pp}(\bm{r}_1, \bm{r}_2, \bm{r}_1', \bm{r}_2') = -G \delta(\bm{R}-\bm{R}')P(\bm{r})P(\bm{r}')\frac{1}{2} (1-P^{\sigma}),
\end{equation}
where $\bm{R}$ is the centre-of-mass coordinate ($\bm{R} = \frac{1}{2}(\bm{r}_1+\bm{r}_2)$), $\bm{r}$ is the relative coordinate ($\bm{r} = \bm{r}_1-\bm{r}_2$), and $P(\bm{r})$ is defined as
\begin{equation}
     P(\bm{r}) = \frac{1}{(4\pi a^2)^{3/2}}\text{e}^{-\frac{r^2}{4a^2}},
\end{equation}
and the parameters $G$ and $a$ are found from the D1S parametrisation of the Gogny force as $G_{p(n)} = 728$ MeV fm$^3$ and $a=0.644$ fm for the DD-ME2 interaction. 
The free energy of the nucleus is given by
\begin{equation}
    F = E - TS,
\end{equation}
where $E$ is the total energy, and $S$ is the entropy;
\begin{equation}
    S = -k_B \sum_k [f_k \text{ln}f_k + (1-f_k)\text{ln}(1-f_k)].
\end{equation}

In this study, we employ the multidimensionally-constrained RHB model to study localisation and clustering in \(\textsuperscript{16}\text{O}\) at finite temperature. The calculations are performed by imposing constraints on the multipole operators and radii of $^{16}\text{O}$. The quadratic constraint method involves variation of the function, subject to quadratic constraints:
\begin{equation}
 \langle \hat{H} \rangle + \sum_{\lambda \mu}  C_{\lambda \mu} (\langle \hat{Q}_{\lambda \mu} \rangle-q_{\lambda \mu})^2,
    \label{eqn:augmented_lagrangian}
\end{equation}
 where $\langle \hat{H} \rangle$ is the total energy, and $\langle \hat{Q}_{\lambda \mu} \rangle$ is the expectation value of the associated mass multipole operator, $\hat{Q}_{\lambda \mu}=r^{\lambda}Y_{\lambda \mu}$, where $r$ is the radius, and $Y_{\lambda\mu}$ are the spherical harmonics. Additionally, $q_{\lambda \mu}$ represents the constrained value of the multipole moment, and $C_{\lambda \mu}$ is the corresponding stiffness constant. Convergence of the results obtained with the quadratic constraint method strongly depend on the magnitude of the stiffness constant. This issue is addressed by implementing the augmented Lagrangian method (see Refs. \cite{Staszczak2010, dirhb}). Also, there must be an additional constraint on $\hat{Q}_{10}=z$ to prevent any spurious motion of the centre of mass, which is given by $\langle Q_{10} \rangle =0$ fm. The dimensionless deformation parameters read
\begin{equation}
    \beta_{\lambda \mu}=\frac{4\pi}{3NR^{\lambda}} \langle \hat{Q}_{\lambda \mu} \rangle
    \end{equation}
where $R=r_{0}A^{1/3}$ with $r_{0}=1.2$ fm and $N$ represents the number of neutrons, protons or nucleons. 
It is important to note that neutron vapour effects \cite{BONCHE1984278, expanding-limits-nuc-stab, PhysRevC.109.014318}, which may have the potential to impact higher temperatures ($T \geq 2$ MeV), have not been accounted for in this study.



\section{\label{sect:results}Results}

\begin{figure}[htbp!]
\centering
\includegraphics[width=\linewidth]{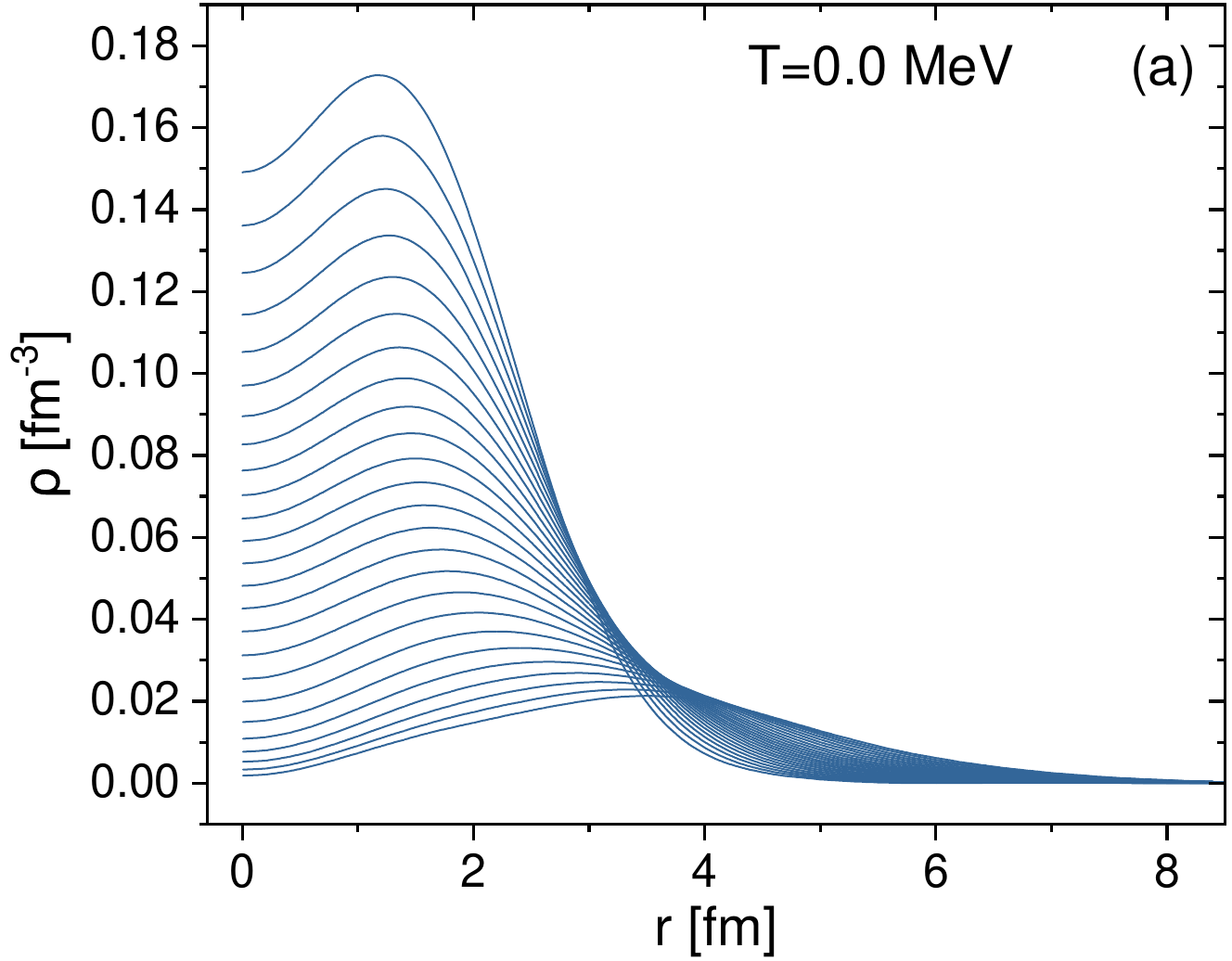}
\includegraphics[width=\linewidth]{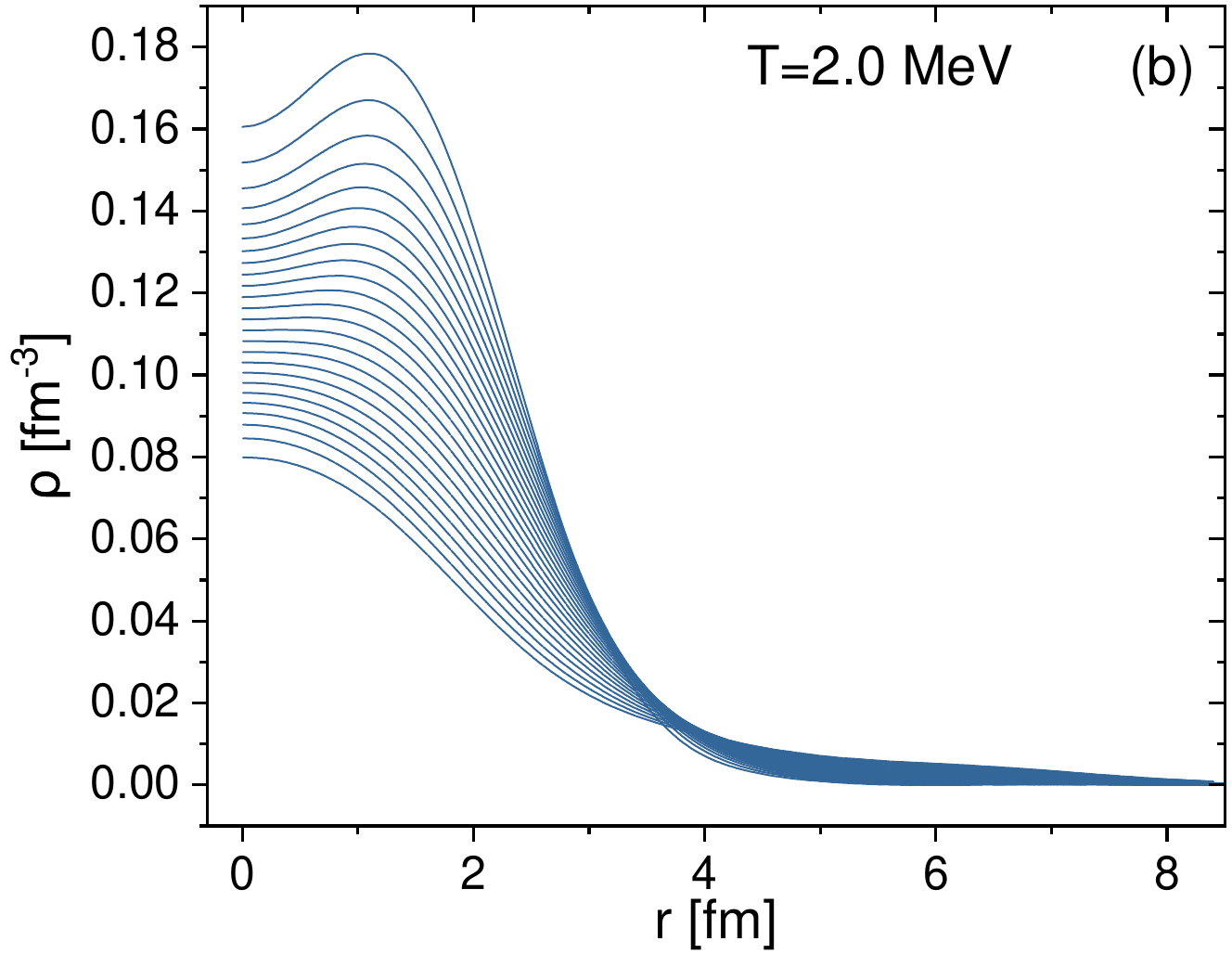}
\caption{Nuclear radial density for $^{16}\text{O}$ at temperatures (a) $T = 0.0$ MeV, and (b) $T = 2.0$ MeV, performed using the multi-constrained FT-RHB. The nucleus undergoes isotropic expansion ($\beta_{2\mu}, \beta_{3\mu} = 0$). The radius is constrained from $R_{\text{rms}} = 2.60$ fm to $R_{\text{rms}} = 5.00$ fm, with steps of 0.10 fm.}
\label{fig:radvar-4}
\end{figure}

It is known that clustering can occur in light nuclei \cite{RevModPhys.90.035004, PhysRevC.87.044307, PhysRevC.90.054329} and dilute nuclear systems \cite{GIROD-clustering, PRC-16O_QPT, PhysRevC.81.015803, doi:10.1126/science.abe4688, REN2024138463}. 
Previously, the formation of $\alpha$ particles or clustering in finite nuclei has been studied by constraining the nuclear radius to larger values, i.e., lower nuclear densities, where (at sub-saturation densities) the dominant order is an $\alpha$-condensate phase, and the associated structural change is driven by nuclear density as the control parameter ~\cite{GIROD-clustering, PRC-16O_QPT}.
In this section, we discuss this transition of nuclei from a homogeneous (delocalised) system to a clustered (localised) system at finite temperatures. To this end, we performed calculations on $^{16}\text{O}$ at both zero and finite temperatures by constraining its radius to larger values and continuously diluting the system. Since we are dealing with a dilute nucleus with a large constrained root mean squared (rms) radius, $R_{rms}$, in a harmonic oscillator (HO) basis, we need to optimise the number of HO shells and associated parameters to avoid unphysical solutions. A smooth decrease of the density distribution is required, as an unphysical solution results in a dense core surrounded by a dilute nucleon cloud \cite{PRC-16O_QPT}, with a physical solution resulting in inflation of the entire nucleus. It was found that using 12 HO shells with $\hbar \omega = 13$ MeV ensured convergence of the calculations, minimised the energy (while retaining the physical nature of the solutions), and maintained a linear decrease in the density distribution. Fig.~\ref{fig:radvar-4} shows the radial density distribution results for $^{16}\text{O}$, obtained by increasing the nuclear radius while constraining all multipolar mass moments to zero, i.e., $Q_{\lambda\mu} = 0$ at $T = 0.0$ MeV and $T = 2.0$ MeV. At zero temperature, the ground-state rms radius of $^{16}\text{O}$ is found to be $R_{gs} = 2.59$ fm. In these panels, the radius is varied from $2.60$ to $5.00$ fm in steps of $\Delta R_{\text{rms}} = 0.10$ fm. The results clearly show a regular decrease in the density of the $^{16}\text{O}$ nucleus, becoming more diffuse under increasing radial constraints at both zero and finite temperatures. This demonstrates that we avoid the unphysical undesired dense nucleus plus neutron cloud solutions.

\begin{figure}[ht!]
    \centering
    \includegraphics[width=\linewidth]{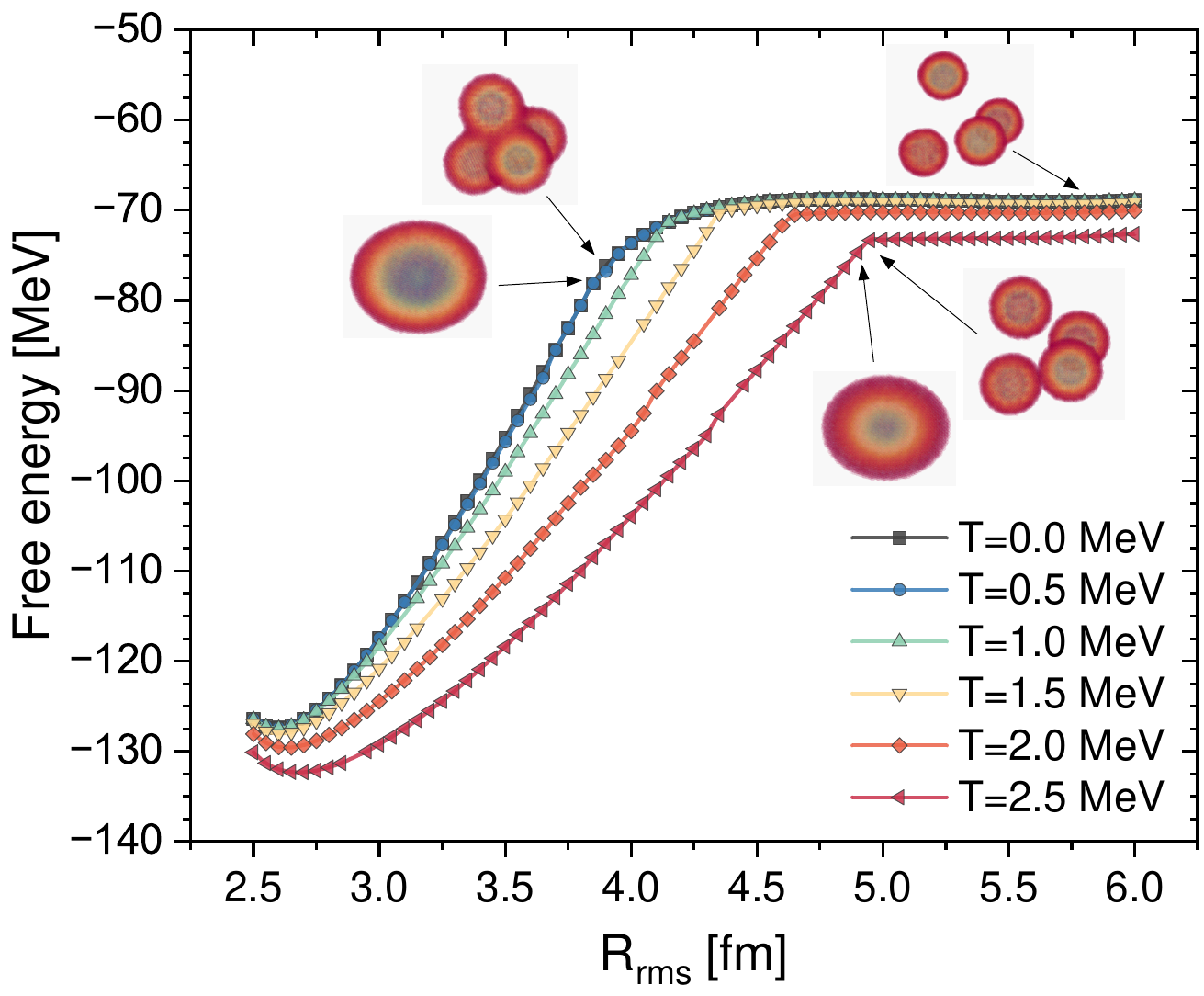}
    \caption{The free energy in $^{16}\text{O}$ as a function of constrained radii for increasing temperature. The calculations are performed using the multi-constrained FT-RHB method with the DD-ME2 interaction. Three-dimensional density ($\rho \geq 0.01 \text{ fm}^{-3}$) plots highlight the emergence of clustering.}
    \label{fig:free-tot}
\end{figure}

Fig.~\ref{fig:free-tot} shows the free energy as a function of the constrained radius for finite temperatures in the range $T = 0.0$ to $2.5$ MeV. The calculations are performed using the multi-constrained FT-RHB method and the DD-ME2 interaction. The nucleus is diluted while simultaneously imposing isotropic inflation through the constraint $Q_{2\mu} = 0$ ($\beta_{2\mu} = 0$). This ensures that, whether the system is homogeneous or clustered, it remains globally spherical. The other multipolar mass moments ($Q_{\lambda\mu}$) are unconstrained, meaning that nuclei can develop axial octupole deformation ($\beta_{30}$) or non-axial octupole (tetrahedral) deformation ($\beta_{32}$) if it is energetically favourable.

At zero temperature, the free energy curve exhibits smooth behaviour: the free energy increases as the nucleus becomes more diluted, and the nucleus favours a tetrahedral configuration ($\beta_{32} \neq 0$) at a critical radius of $R_{\text{c}} = 3.90$~fm, corresponding to $\rho_{\mathrm{Mott}}/\rho_0 = (R_{gs}/R_{\text{c}})^3 \approx 0.3$. This result is also in agreement with previous studies \cite{GIROD-clustering, PRC-16O_QPT}. At this point, the nucleus undergoes an abrupt transition from a spherical configuration to a four-$\alpha$ configuration. In this way, the nucleus forms localised clusters that restore the saturation density and stabilise the free energy, instead of remaining in a dilute homogeneous phase and becoming unbound. Within the plot, we also display nuclei at selected radii in a 3-dimensional density representation. The transition between the two distinct phases is highlighted by showing nuclei under sequential radial constraints: the system initially exhibits a homogeneous density distribution, which evolves into a clustered tetrahedral configuration beyond the critical point. By further diluting the nucleus, a tetrahedron of four $\alpha$-particles becomes more apparent and move further apart, while the development of this tetrahedral configuration stabilises the free energy curve up to $R_{\text{rms}} = 6.00$ fm. 
By increasing the temperature, we observe a similar behaviour, namely that the nucleus undergoes a transition from a spherical configuration to a four-$\alpha$ configuration. At $T = 0.0$ and $T = 0.5$ MeV, the free energy curve shows little variation, and the results are essentially overlapping. The first notable effect of the temperature is observed at $T=1.0$ MeV,  where the free energy curve shifts downward, and the clustering occurs at larger radii. As the temperature increases, the impact becomes more pronounced, and the free energy curves continue to shift downward and show a more distinct transition compared to the zero-temperature case.

The formation of a tetrahedral shape at larger radii is also explained in terms of the single-particle levels, as discussed in Ref.~\cite{PRC-16O_QPT}. It is well known that the ground state of $^{16}\text{O}$ is spherical due to the $p$-shell closures. Isotropic inflation leads to a reduction of the Fermi gap, causing $^{16}\text{O}$ to become a nearly degenerate system \cite{PRC-16O_QPT}. The system will therefore seek to remove this degeneracy and rearrange itself. This can be done through either the development of pairing correlations or angular correlations (in the form of axial or non-axial octupole deformed configurations), depending on which is more energetically favourable. In this work, the development of pairing correlations was not observed. It was previously anticipated that the development of deformation within the nucleus would lead to a more energetically favourable system \cite{PRC-16O_QPT}, and our calculations confirm this expectation.

It is important to note that the constrained RHB framework includes mean-field correlations through the self-consistent interaction of nucleons with meson fields, along with static pairing (two-body) correlations via the Bogoliubov transformation and Pauli correlations. However, it does not account for more complex many-body effects, such as thermal shape fluctuations~\cite{PhysRevC.106.054309} and restoration of broken symmetries~\cite{ZHOU2016227,PhysRevC.97.024334}, which may influence the precise location and character of the transition. Nevertheless, the FT-RHB model captures spontaneous symmetry breaking, which is essential for cluster localisation, and provides a consistent mean-field signature of the structural transition.

The development of $\beta_{32}$ is important for signalling a clustered state within a dilute nucleus. It is worth clarifying that for radii near the ground-state radius, a finite non-axial octupole deformation parameter would simply result in a tetrahedrally deformed nucleus with nucleons delocalised within a small volume—namely, without clustering. However, below the critical Mott density, it becomes favourable to form clusters. The configuration of the four $\alpha$-particles, while minimising the energy, adopts a tetrahedral shape with a non-zero $\beta_{32}$ value.

\begin{figure}[htbp!]
    \centering
    \includegraphics[width=\linewidth]{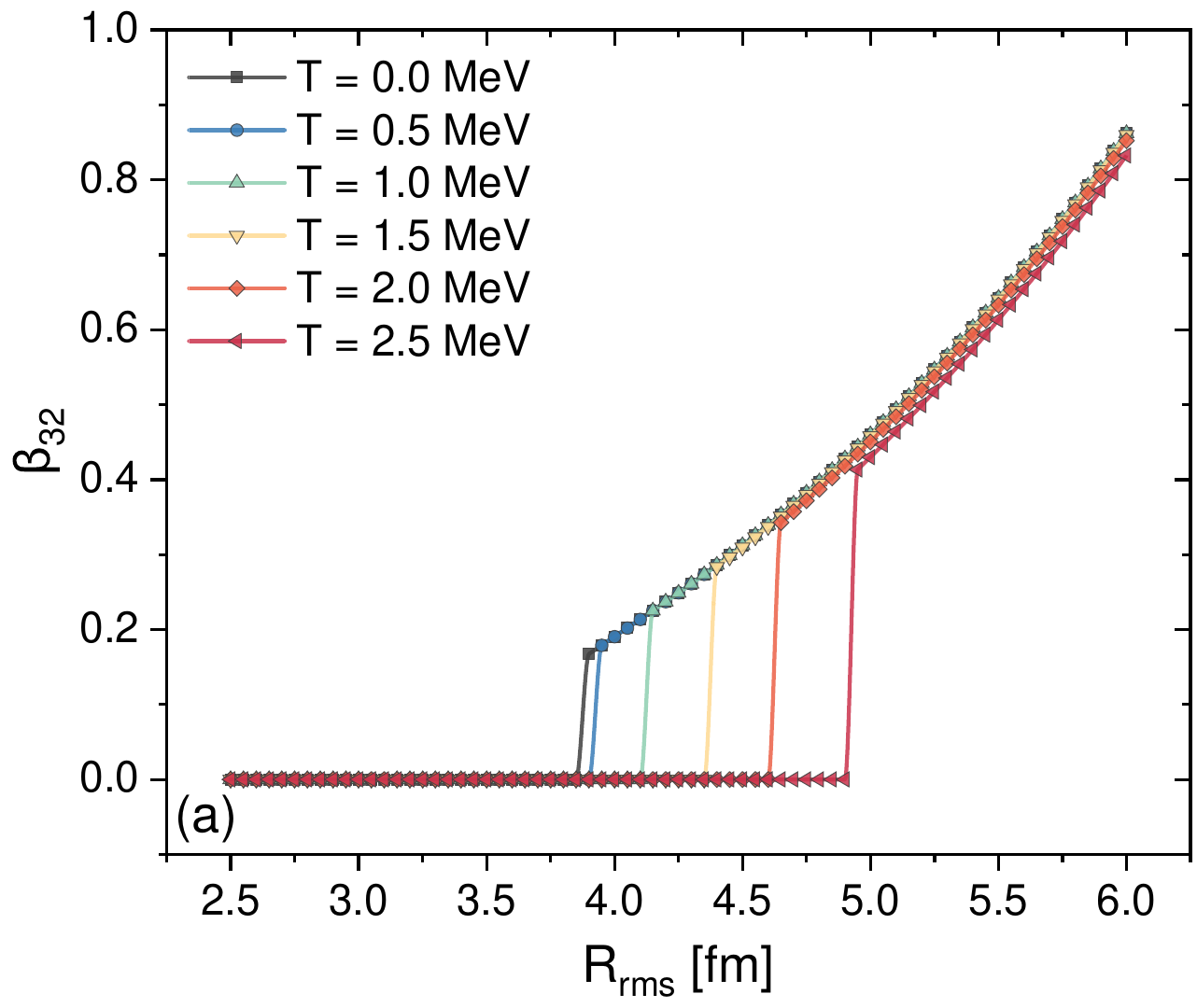}
    \includegraphics[width=\linewidth]{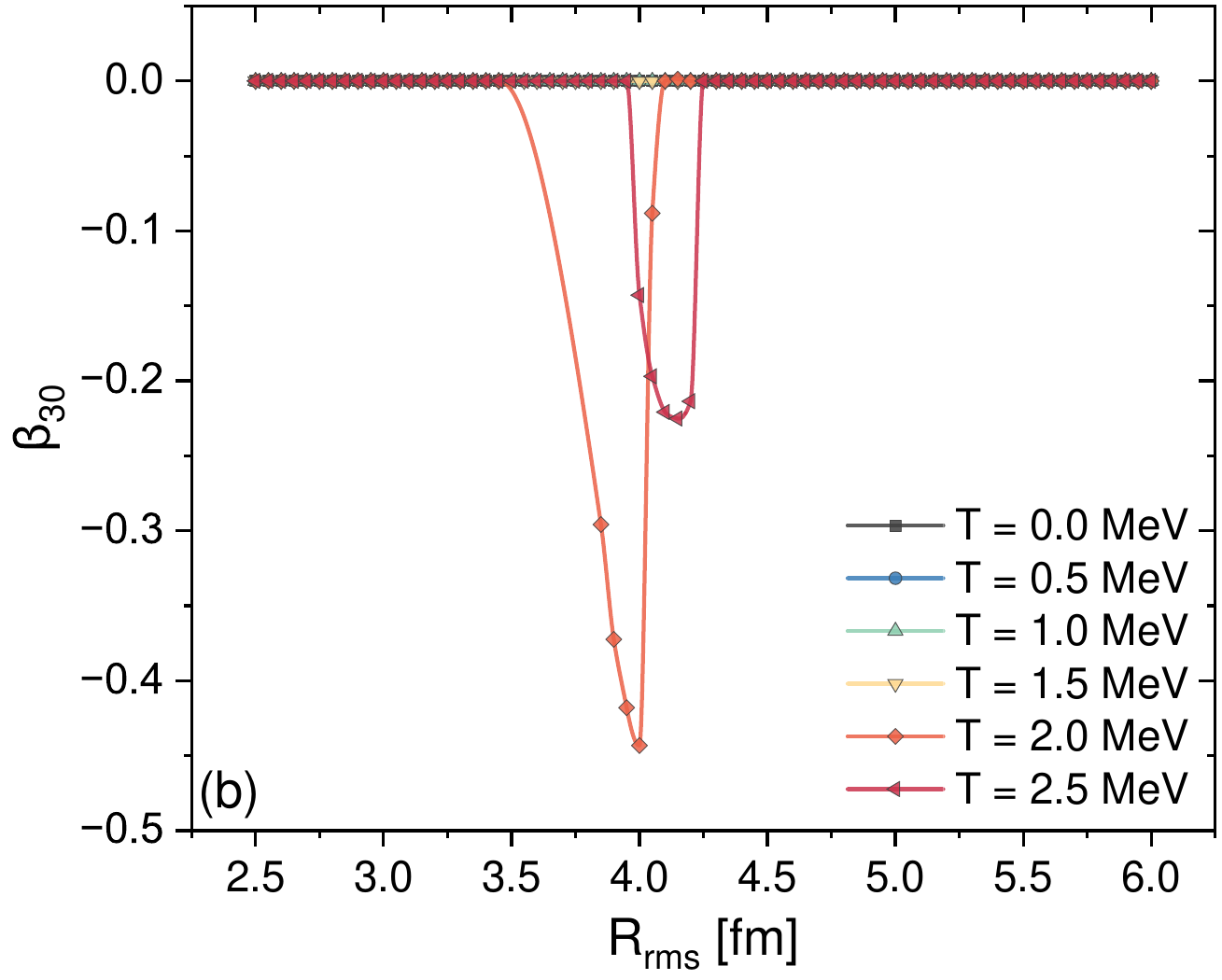}
    \caption{The evolution of (a) the non-axial octupole deformation, $\beta_{32}$, and (b) the octupole deformation, $\beta_{30}$ in $^{16}\text{O}$, under increasing radial constraint and temperature.}
    \label{fig:b30-b32}
\end{figure}

Figure~\ref{fig:b30-b32}(a) illustrates the evolution of the non-axial octupole deformation $\beta_{32}$ as a function of the constrained radius $R_{\text{rms}}$ at finite temperatures, in relation to the energy curves shown in Fig.~\ref{fig:free-tot}. Firstly, it is evident that (once it has developed), the magnitude of the non-axial octupole deformation increases with increasing constrained radius, indicating that the nucleus requires a larger tetrahedral configuration to become stabilised as the nucleus becomes more inflated. Secondly, the figure highlights the strong correlation between the emergence of a non-zero $\beta_{32}$ and the clustering. Specifically, the onset of a non-zero $\beta_{32}$ signifies the transition from a homogeneous to a clustered configuration within the dilute nucleus. 
This correlation is further confirmed through the analysis of 3-dimensional density distributions. 
As mentioned above, the transition from a homogeneous to a localised configuration is delayed with increasing temperature, however once localisation develops, the magnitude of the non-axial octupole deformation remains almost the same across all temperatures. The delay in the cluster formation as the temperature increases is related to the impact of the temperature on the shell structure of nuclei. It is known that at finite temperatures the nucleus gets additional excitation energy from the environment, and this leads to changes in the population of single-particle energy levels in the vicinity of the Fermi surface. 
At high temperatures, particularly for $T > 1$ MeV, the shell effects vanish, and the depopulation of intruder states—responsible for driving deformation—causes deformed nuclei to adopt a spherical shape. In other words, for deformed nuclei, temperature leads to the vanishing of deformation properties at critical temperatures, and the nucleus becomes spherical \cite{PhysRevC.34.1942, GOODMAN198130, PhysRevLett.85.26, PhysRevC.93.024321, PhysRevC.109.014318}. In this study, $^{16}$O is inflated by constraining its radius at finite temperatures, and the development of a tetrahedral configuration is delayed as the temperature increases due to the presence of additional excitation energy in the system, which prevents the formation of a tetrahedral configuration for a while.


Additionally, the system is free to develop axial octupole deformation. Figure \ref{fig:b30-b32}(b) illustrates the development of axial octupole deformation, $\beta_{30}$, during radial expansion at different temperatures. At lower temperatures, $\beta_{30}$ does not develop, regardless of the radius. However, at higher temperatures, i.e., at $T = 2.0, 2.5$ MeV, $\beta_{30}$ develops prior to the cluster formation. It is important to note that an investigation of the 3-dimensional densities reveals no localisation with only $\beta_{30}\neq 0$, and the nucleus remains homogeneous despite the development of this octupole deformation. 

\begin{figure}[htbp!]
    \centering
    \includegraphics[width=\linewidth]{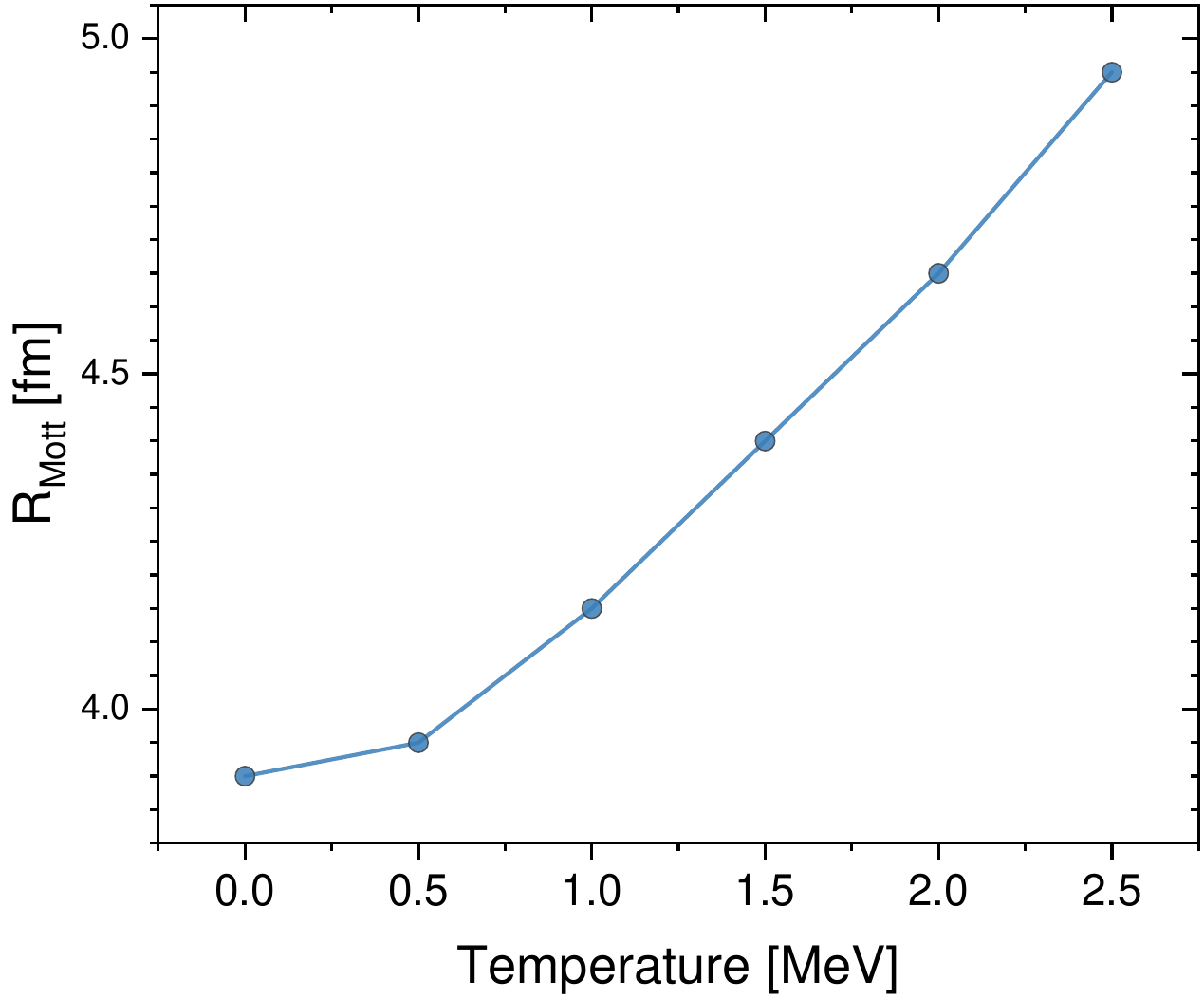}
    \caption{The relationship between the critical (Mott) radius, at which clustering occurs, and the temperature (see text for details).}
    \label{fig:Rcrit-Temp}
\end{figure}

As mentioned above, the onset of clustering is delayed due to the competing effects of deformation and temperature at finite temperatures. This relationship between the critical radius $R_{\text{c}}=R_{\text{Mott}}$ for the clustering and temperature is shown in Fig.~\ref{fig:Rcrit-Temp}, where a linear relationship is observed for temperatures above $T = 0.5$ MeV. More explicitly, at a temperature of $T = 0.5$ MeV, the cluster formation occurs with the development of $\beta_{32} \neq 0$, observed at an rms radius of $R_{\text{Mott}}= 3.95$ fm, which is slightly higher but still close to the value at zero temperature. This is observed in Fig.~\ref{fig:Rcrit-Temp}, where a much shallower gradient indicates that temperature and thermal effects are not yet significant. By increasing temperature further, at $T = 1.0$ MeV, localisation occurs at $R_{\text{Mott}} = 4.15$ fm, and at $T = 1.5$ MeV, it occurs at $R_{\text{Mott}} = 4.40$ fm. As the temperature increases, the onset of cluster formation gradually shifts to larger radii, indicating that higher thermal energy delays the development of $\alpha$-cluster structures. This trend continues up to $T = 2.0$ MeV and $T = 2.5$ MeV, where transitions occur at $R_{\text{Mott}} = 4.65$ fm and $R_{\text{Mott}} = 4.95$ fm, respectively. 

\begin{figure}[htbp!]
    \centering
    \includegraphics[width=\linewidth]{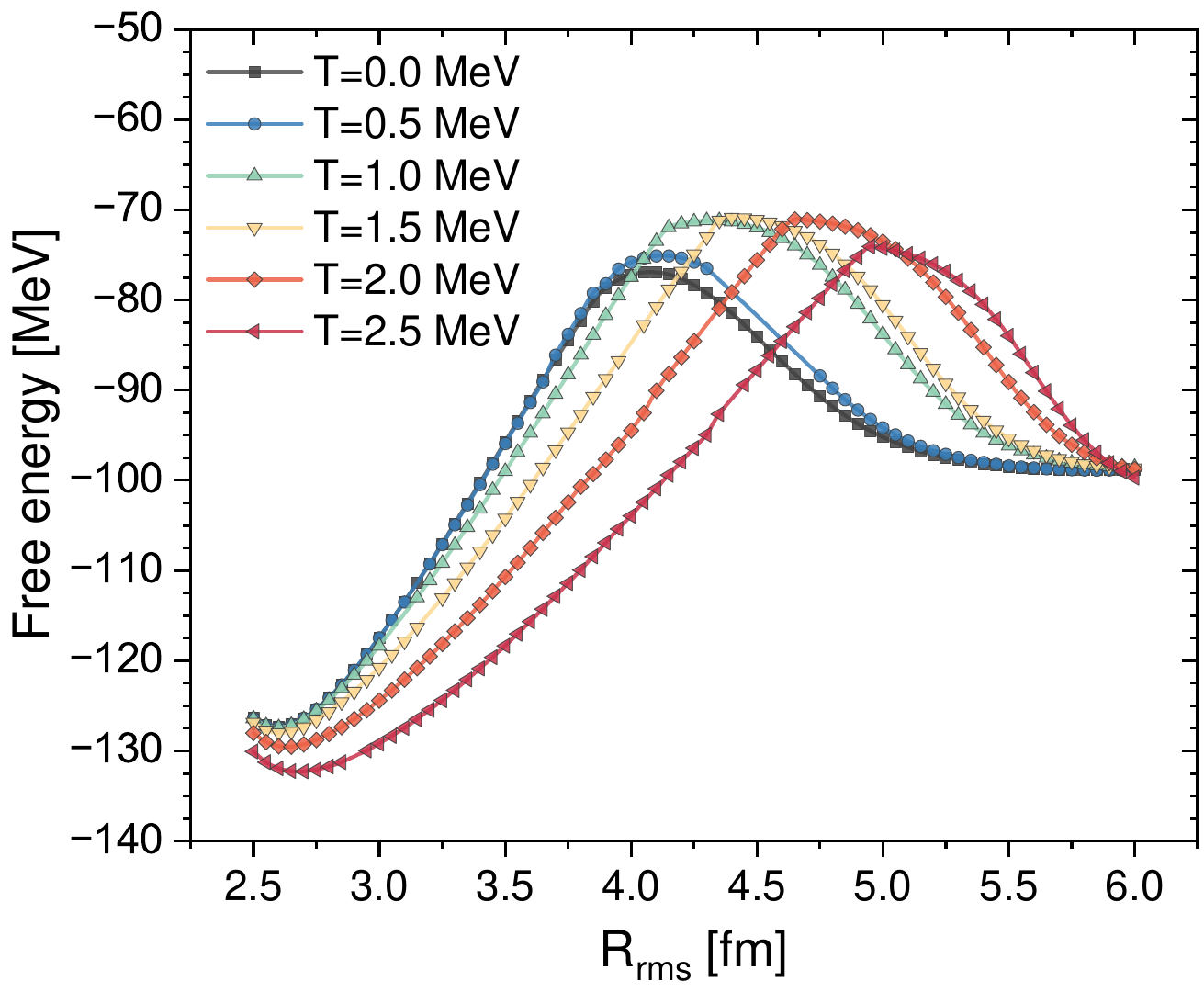}
    \caption{The same as in Fig.~\ref{fig:free-tot}, but with the zero-point energy (ZPE) correction applied for the free energy of the system with $\alpha$-particles.}
    \label{fig:free-ZPE-corr}
\end{figure}

It is also important to acknowledge that once $\alpha$-cluster formation occurs, the $\alpha$'s carry their own spurious centre-of-mass energy, which must be accounted for and eliminated. However, there is no simple microscopic way to achieve this. In Ref. \cite{GIROD-clustering}, this was done by removing $\approx$ 7 MeV for each $\alpha$-particle from the system. In this work, a Gaussian function is folded into the theoretical results to remove the zero-point energy (ZPE) of the $\alpha$ particles. The assumption is that the ZPE is zero for a homogeneous nucleus, and $30$ MeV for well-separated $\alpha$ particles, and takes the following form:

\begin{equation}
    \text{ZPE}_{\text{corr}}(R_\text{rms}) = \frac{30}{1 + \exp(-\sigma(R_{\text{rms}} - \eta))},
    \label{eqn:ZPE-correction}
\end{equation}
where $\sigma$ and $\eta$ are temperature-dependent parameters, adjusted for each temperature and given in Table~\ref{tab:sigma_eta_values}. The temperature dependence of the ZPE arises through parameters that reflect the spatial configuration and relative distances between the $\alpha$ clusters, which evolve with temperature due to thermal expansion and delocalisation. These parameters are extracted from the intrinsic density distributions obtained self-consistently at each temperature within our constrained RHB calculations. The ZPE correction to the free energy is shown in Fig. \ref{fig:free-ZPE-corr}. A 30 MeV difference can be seen near $R_{\text{rms}} = 6.00$ fm, which corresponds to the radial expansion where the $\alpha$-particles are considered well-separated.  

\begin{table}[htb!]
\caption{{\label{tab:sigma_eta_values} Temperature-dependent values of the parameters $\sigma$ and $\eta$ used in the ZPE correction to the free energy.}}
\begin{ruledtabular}
{\begin{tabular}{ccccc}
 & $T$ [MeV] & $\sigma$ [fm$^{-1}$] & $\eta$ [fm] & \\
\hline
 & 0.0 & 4.00 & 4.50 & \\
 & 0.5 & 4.25 & 4.60 & \\
 & 1.0 & 4.50 & 5.00 & \\
 & 1.5 & 4.75 & 5.10 & \\
 & 2.0 & 5.20 & 5.40 & \\
 & 2.5 & 5.60 & 5.60 & \\
\end{tabular}}
\end{ruledtabular}
\end{table}

\begin{figure}[htb!]
    \centering
    \includegraphics[width=\linewidth]{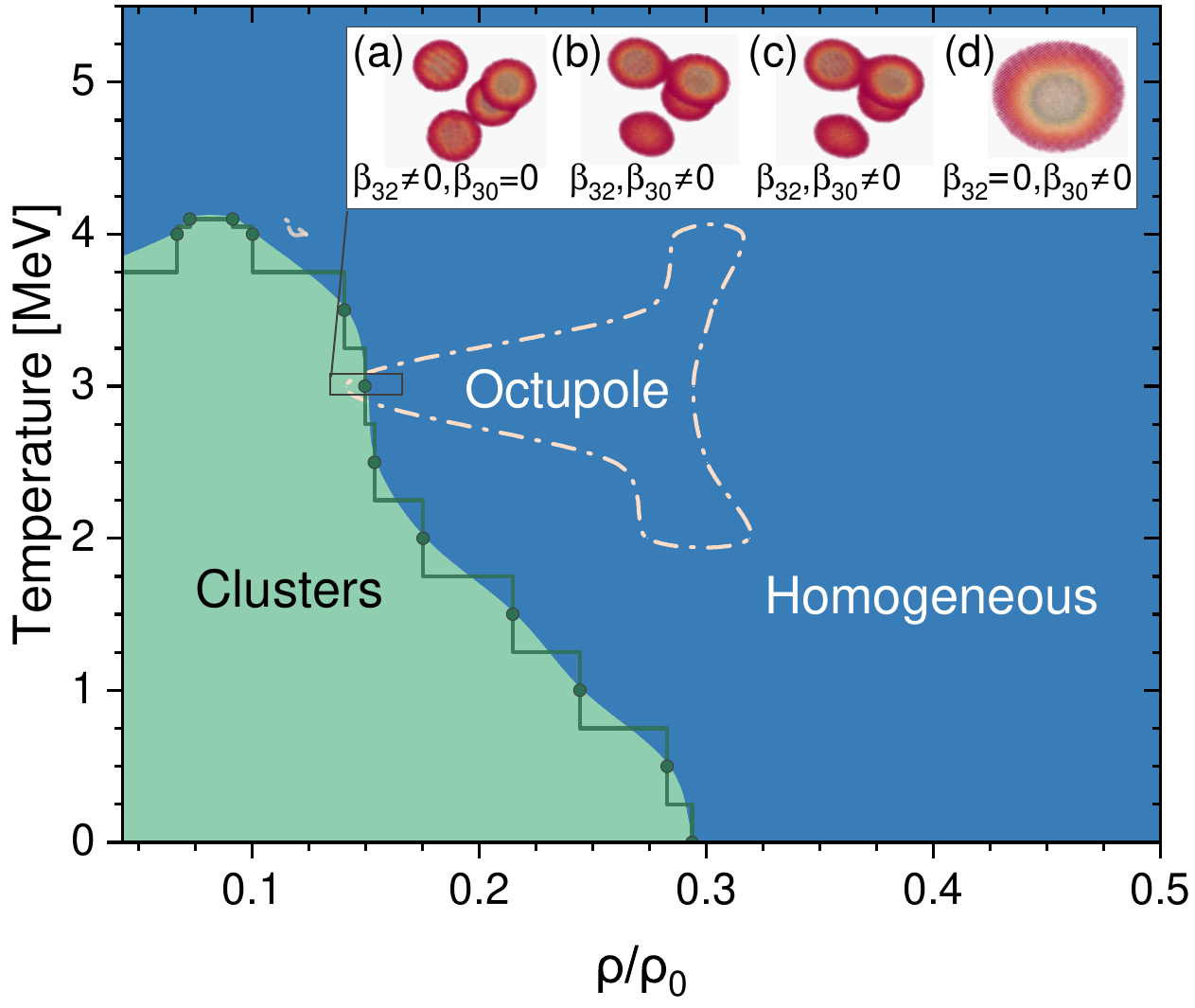}
    \caption{The regions of nuclear homogeneity and clustering in $^{16}\text{O}$ as functions of temperature and normalised density are shown. The regions with non-zero octupole moments are also highlighted, although they do not induce localisation. The inserted panel highlights this and shows the three-dimensional densities ($\rho \geq 0.01 \text{ fm}^{-3}$) at $T = 3.0$ MeV around the transition for $R_{\text{rms}}$: (a) 5.30 fm, (b) 5.25 fm, (c) 5.20 fm, and (d) 5.15 fm.}
    \label{fig:HC-rhomottrho}
\end{figure}

In Fig.~\ref{fig:HC-rhomottrho}, we also present a temperature-density plane, which shows the regions of homogeneous and clustered configurations for $^{16}\text{O}$. The temperature is shown as a function of the normalised density, and the boundary between these phases highlights the critical density ($\rho_{\mathrm{Mott}}/\rho_0$) and temperature at which clustering occurs. The solid \textit{Clusters-Homogenous} region is formed by smoothing the results between data points, with the line around the boundary representing the results of the calculations using a discrete step function, where the step size for $R_{\text{rms}}$ and temperature is $0.05$ fm and 0.5 MeV, respectively. The relationship between the normalised density and radius at the boundary is through the simple relationship $\rho_{\mathrm{Mott}}/\rho_0 = (R_{gs}/R_{\text{c}})^3$. Higher temperatures are also included to highlight the point at which clustering no longer occurs. This also allows us to further explore the competition between $\alpha$-particle clustering and the loss of deformation due to the increase in temperature.

As mentioned above, at zero temperature, cluster formation occurs at a normalised density of $\rho_{\mathrm{Mott}}/\rho_0 \approx 0.3$, which is consistent with the predictions of previous studies \cite{PRC-16O_QPT, GIROD-clustering}. This is also comparable to the density of the Hoyle state, which, given that the rms radius is approximately 1.5 times larger than the ground state radius \cite{hoyle_spatially_extended}, corresponds to $\rho_{\mathrm{Mott}} \approx \rho_0/4 - \rho_0/3$. As the temperature increases, the onset of clustering shifts to progressively lower normalised densities, until the dilution becomes so significant that the nucleus no longer favours the emergence of localised structures. The highest temperature at which cluster emergence is observed is $T = 4.1$ MeV, and it occurs at $\rho_{\mathrm{Mott}}/\rho_0 \approx 0.09$. At temperatures higher than this, clustering no longer emerges. The islands of non-zero octupole deformation are also highlighted for completeness, though they do not induce localisation. As noted earlier, non-zero octupole deformation can occur before or after the cluster formation. This is illustrated in the inset panel of Fig.~\ref{fig:HC-rhomottrho}, which displays the 3-dimensional density distributions (\(\rho \geq 0.01~\text{fm}^{-3}\)) of the nucleus at increasing normalised densities—or equivalently, decreasing radii—around the transition point at \(T = 3.0\)~MeV.
 Fig.~\ref{fig:HC-rhomottrho}(a) shows a pure tetrahedral shape with non-zero $\beta_{32}$ but no $\beta_{30}$ deformation at $R_{rms}=5.30$ fm, as found at the transition for all other temperatures. Examining the density plots in Fig.~\ref{fig:HC-rhomottrho}(b) and (c) -- at $R_{rms}=5.25$ fm and $R_{rms}=5.20$ fm, respectively -- have both non-zero $\beta_{32}$ and $\beta_{30}$, and a non-uniform tetrahedral structure is observed, in which two of the alpha clusters are not as spatially separated, resembling a more $^{8}\mathrm{Be}\text{-}2\alpha$ configuration. Although a non-zero $\beta_{30}$ can cause distortion of the uniform tetrahedral structure (when present alongside a non-zero $\beta_{32}$), a non-zero $\beta_{30}$ in isolation does not induce localisation, as demonstrated by the inset in Fig.~\ref{fig:HC-rhomottrho}(d). A smaller island of octupole deformation also appears at $T = 4.1$ MeV at a much lower normalised density compared to the rest of the octupole deformation development, but it still occurs before cluster formation and does not induce localisation. It should be reminded to the reader that although the neutron vapour can be important at high temperatures \cite{expanding-limits-nuc-stab, PhysRevC.109.014318}, it is not included in these calculations and will be addressed in future work.

We also compare our findings with previous studies for symmetric nuclear matter at finite temperatures. It is evident that making a one-to-one comparison is not feasible, as finite nuclei and nuclear matter are distinct systems. Although both exhibit clustering under certain conditions, the underlying physical mechanisms and temperature ranges differ significantly. Nonetheless, we can make at least qualitative comparisons. In Ref. \cite{PhysRevC.81.015803}, the authors study the $\alpha$-particle fraction in symmetric nuclear matter at temperatures between $T = 4$ and $20$ MeV using the generalised relativistic mean-field (RMF) model. They found that at lower temperatures, the particle fraction increases, making $\alpha$-particles more stable and prevalent, particularly at low densities. Higher temperatures reduce the particle fraction due to increased thermal motion, which disrupts $\alpha$-particle formation. Our results display similar behaviour and are consistent with the findings in Ref.~\cite{PhysRevC.81.015803}, although our temperature range is lower, as we deal with finite nuclei and shell effects are significant.
At low densities ($\rho \lesssim \rho_0/3$) and low temperatures, $\alpha$-particles are more prevalent and stable. However, at higher densities and temperatures, $\alpha$-particles become less stable and dissociate. The concept of $\alpha$-particle condensation in low-density infinite matter, as defined by the Thouless criterion in Ref.~\cite{PhysRevLett.80.3177}, identifies the transition point at which $2\mu_n + 2\mu_p$ equals the $\alpha$ binding energy. While this framework is appropriate for nuclear matter with well-defined chemical potentials and explicit $\alpha$ degrees of freedom, it cannot be directly applied in our FT-RHB model of finite nuclei due to the lack of well-defined asymptotic particle numbers and the absence of explicit $\alpha$ degrees of freedom in the mean-field description. Nevertheless, our findings in Fig. \ref{fig:HC-rhomottrho} also suggest qualitative agreement with Fig. 2 in Ref.~\cite{PhysRevLett.80.3177}.

Recent experimental studies, such as Ref.~\cite{CHEN20231119}, investigate the Hoyle-like structure in ${}^{16}$O through inelastic scattering. The findings reveal four narrow resonances above the 15.1 MeV state, providing compelling evidence for the existence of high-lying four-$\alpha$ resonant states in ${}^{16}$O. In Ref.~\cite{sym13091562}, the authors analyse experimental data from heavy-ion collisions to study $\alpha$-clustering in excited self-conjugate nuclei. They measured the excitation energy and determine the temperature using Maxwellian fits, while the density conditions required for clustering are inferred from simulations. The mean value of the excitation energy, temperature, and density normalised to saturation density for \(^{16}\)O are obtained as \(\langle E^* \rangle = 52.4 \pm 15.7\)~MeV, \(T = 6.15 \pm 0.03\)~MeV, and \(\rho/\rho_0 = 0.37 \pm 0.04\), respectively. To facilitate comparison with these results, we calculate the excitation energy associated with the transition at various temperatures using
\[
E^* = E(T, R_{\text{Mott}}) - E(T=0, R_{\text{gs}}),
\]
where the first term represents the total energy of the nucleus at finite temperature at the onset of clustering, and the second term corresponds to the total energy of the ground state at zero temperature.
The results are given in Table~\ref{tab:ex_energies}, along with the associated normalised Mott density. As can be seen, we also obtain similar excitation energies; however, they occur at much lower temperatures and densities since the nucleus is also diluted by constraining its radii to large values in our calculations. It is important to note that the excited cluster states observed in recent experiments~\cite{sym13091562,CHEN20231119} correspond to specific, quantised configurations of nucleons and do not represent the full thermodynamic behaviour of nuclear systems at finite temperature. In contrast, our theoretical framework describes a finite nucleus in thermal equilibrium using the grand canonical ensemble, where thermal effects are incorporated via temperature-dependent occupation probabilities of quasiparticle states. As a result, the model provides averaged properties of the system rather than discrete excitation spectra. While a direct one-to-one comparison with individual excited states is not possible, our results offer complementary insights into the conditions under which cluster-like structures become energetically favourable in finite nuclei.

\begin{table}[htb!]
\caption{\label{tab:ex_energies}  Temperature (\(T\)), excitation energy (\(E^*\)), and normalised Mott density (\(\rho_{\text{Mott}}/\rho_0\)) for \(^{16}\)O at the structural transition point.}
\begin{ruledtabular}
\begin{tabular}{ccccc}
 & $T$ [MeV] & $E^*$ [MeV] & $\rho_{Mott}/\rho_0$ & \\
\hline
 & 0.5 & 52.6 & 0.28 & \\
 & 1.0 & 56.4 & 0.24 & \\
 & 1.5 & 59.9 & 0.21 & \\
 & 2.0 & 65.6 & 0.18 & \\
 & 2.5 & 76.4 & 0.16 & \\
\end{tabular}
\end{ruledtabular}
\end{table}

Finally, it is also worth noting that all the calculations presented here are performed with the DD-ME2 interaction. Calculations using the DD-PC1 interaction have also been conducted, yielding a similar clustering behaviour during the inflation of the \(^{16}\text{O}\) nucleus. Although the exact location of \(R_{\text{c}}\), where the cluster formation occurs, depends on the type of interaction used, the overall behaviour—which is the key focus of this investigation—remains unchanged.

\section{\label{sect:conc}Conclusion }
In this work, we investigated the emergence of $\alpha$-cluster structures in \(^{16}\text{O}\), marking the transition from a homogeneous nuclear system to a localised, clustered configuration. This transition was studied by constraining the nuclear radius and diluting the system at both zero and finite temperatures. To this end, we performed calculations using the multi-constrained FT-RHB with the DD-ME2 interaction. The density of $^{16}\text{O}$ is reduced by increasing the constrained radius, under the condition of isotropic expansion ($\beta_{2\mu}=0$). Then, we explored the effect of temperature on the formation of $\alpha$-clusters on the diluted $^{16}\text{O}$ nucleus.

At zero temperature, we observe a transition at a critical radius from a homogeneous system to a tetrahedral-clustered configuration, corresponding to $\rho_{\text{Mott}} / \rho_0 \approx 0.3$, as indicated by the development of non-axial octupole $\beta_{32}$ deformation. This behaviour remains consistent at finite temperatures; however, with increasing temperature, the the onset of clustering is shifted to larger nuclear radii, corresponding to lower nuclear densities. This happens because of the impact of temperature on the shell structure and deformation properties of nuclei: temperature delays the formation of the tetrahedral configuration and $\alpha$-particle clusters by providing energy to the system. We found that $\alpha$-particle clusters appear up to $T=4.1$ MeV at $\rho_{Mott}/\rho_0 \approx 0.09$ (equivalent to $\rho_{Mott}\approx \rho_0/11$). The development of non-zero octupole $\beta_{30}$ deformation is also possible at finite temperatures, but it does not lead to localisation within the nucleus.

The present study highlights the significance of temperature in the formation of $\alpha$-particle clusters in a diluted nucleus. It is known that neutron vapour may become important at high temperatures, especially for neutron-rich nuclei, and the subtraction of neutron vapour would be necessary. In the forthcoming study, we plan to extend our model to account for this and extend our investigation to neutron-rich nuclei in order to study $\alpha$-clustering at high temperatures.

\section*{Acknowledgments}
E.Y. and P.S. acknowledge support from the UK STFC under award nos. ST/Y000358/1 and ST/V001108/1.

\bibliographystyle{apsrev4-2}
\bibliography{apssamp}

\end{document}